\ifx\pdfminorversion\undefined\else\ifnum\pdfminorversion<7\relax\pdfminorversion=7\relax\fi\fi
\documentclass[aps,prl,twocolumn,showpacs,superscriptaddress,groupedaddress,preprintnumbers]{revtex4-2}

\usepackage{graphicx}
\usepackage{dcolumn}
\usepackage{bm}
\usepackage[english]{babel}
\usepackage[utf8]{inputenc}
\usepackage{float}
\usepackage{hyperref}
\usepackage{amssymb}   
\usepackage{amsmath}   
\usepackage{color}

\begin{document}

\preprint{LA-UR-23-28229,CETUP-2023-001}

\title{Confronting axial-vector form factor from lattice QCD with MINERvA antineutrino-proton data}

\author{Oleksandr Tomalak}
\email{tomalak@lanl.gov}
\author{Rajan Gupta}
\email{rg@lanl.gov}
\author{Tanmoy Bhattacharya}
\email{tanmoy@lanl.gov}
\affiliation{Theoretical Division, Los Alamos National Laboratory, Los Alamos, NM 87545, USA}

\date{\today}

\begin{abstract}
We compare recent MINERvA antineutrino-hydrogen charged-current measurements to phenomenological predictions of the axial-vector form factor based on fits to all available electron scattering and deuterium bubble-chamber data and to representative lattice-QCD (LQCD) determination by the PNDME Collaboration. While there is $1$--$2\sigma$ agreement in the cross section with MINERvA data for each bin in $Q^2$, we identify three regions with different relevance and opportunity for LQCD predictions. For $Q^2 \lesssim 0.2~\mathrm{GeV}^2$, the phenomenological extractions have large number of data points and LQCD is competitive, while MINERvA data have large errors. For $0.2~\mathrm{GeV}^2 \lesssim Q^2 \lesssim 1~\mathrm{GeV}^2$, LQCD is competitive with the MINERvA determination, and both give values larger than from phenomenological extraction. For $Q^2 > 1~\mathrm{GeV}^2$, the MINERvA data are the most precise. Our analysis indicates that with improving precision of MINERvA-like experiments and LQCD data, the uncertainty in the nucleon axial-vector form factor will be steadily reduced.
\end{abstract}

\maketitle

Theoretical and phenomenological predictions of the (anti)neutrino scattering cross sections on nuclear targets such as $^{12}$C, $^{16}$O, and $^{40}$Ar are crucial for understanding results of modern and future neutrino oscillation and cross-section experiments, such as T2K, NOvA, MINERvA, MicroBooNE, SBN, Hyper-K, and DUNE~\cite{Nunokawa:2007qh,NOvA:2007rmc,T2K:2011qtm,KamLAND:2013rgu,MicroBooNE:2015bmn,JUNO:2015zny,Hyper-KamiokandeProto-:2015xww,McConkey:2017dsv,Machado:2019oxb,MicroBooNE:2019nio,Farnese:2019xgw,T2K:2019bcf,NOvA:2019cyt,DUNE:2020ypp}. In this paper we discuss the extraction of the nucleon axial-vector form factor from Lattice QCD, MINERvA experiment, and phenomenological analyses, and provide a comparison between them. It is a key input in current theoretical analyses to predict the cross section using nuclear many-body calculations. Uncertainties in it, together with those from nuclear effects, are the dominant sources of error that need to be reduced~\cite{Ruso:2022qes,Kronfeld:2019nfb}.

The phenomenological nucleon axial-vector form factor, $F_A(Q^2)$, is extracted mainly from the bubble-chamber data collected during the decades of 1970 and 1980~\cite{Mann:1973pr,Barish:1977qk,Miller:1982qi,Baker:1981su,Kitagaki:1983px}, and traditionally parameterized by the dipole form with relatively small uncertainty, see Ref.~\cite{Bernard:2001rs} for a review. The same data were recently reanalyzed using a $z$-expansion fit form for the form factor, and a much more conservative error estimate was obtained~\cite{Meyer:2016oeg}. This latter fit now serves as a phenomenological benchmark for the evaluation of charged-current elastic (anti)neutrino-nucleon scattering cross sections. However, in the extraction of these data, models were used for including nuclear corrections, consequently there may be unquantified systematics as discussed in Ref.~\cite{Meyer:2016oeg}. Looking ahead, there are no approved experiments that would improve these data.

This year, the MINERvA Collaboration presented a novel experimental technique for isolating antineutrino scattering off the hydrogen atoms inside the hydrocarbon molecule, ${\overline \nu}_\mu + p \to n + \mu^+$, and extracted the nucleon axial-vector form factor free from nuclear corrections~\cite{MINERvA:2023avz}. Only events with the angle of the final muon $\theta_\mu \le 20^0$, and momentum $ 1.5~\mathrm{GeV} \le p_\mu \le 20~\mathrm{GeV}$, which are also efficiently measured by the MINOS near{-}detector located just downstream of MINERvA, were selected. Charged-current scattering on hydrogen atoms was separated from the background scattering off carbon nuclei using kinematics. The direction of $\overline \nu$, muon and scattered neutron from hydrogen target defines a plane, whereas a neutron coming from $\overline \nu$ scattering off carbon has a broad distribution about such a plane due to nuclear effects. Using Monte Carlo simulations of these distributions allowed the Collaboration to separate scattering off hydrogen versus carbon. This procedure was cross checked with a beam of neutrinos instead of antineutrinos, for which there are no events with scattering off ``free" hydrogen as the relevant charged-current reaction is ${\nu}_\mu + n \to p + \mu^-$.

\begin{figure*}     
    \centering
    \includegraphics[width=0.58\textwidth]{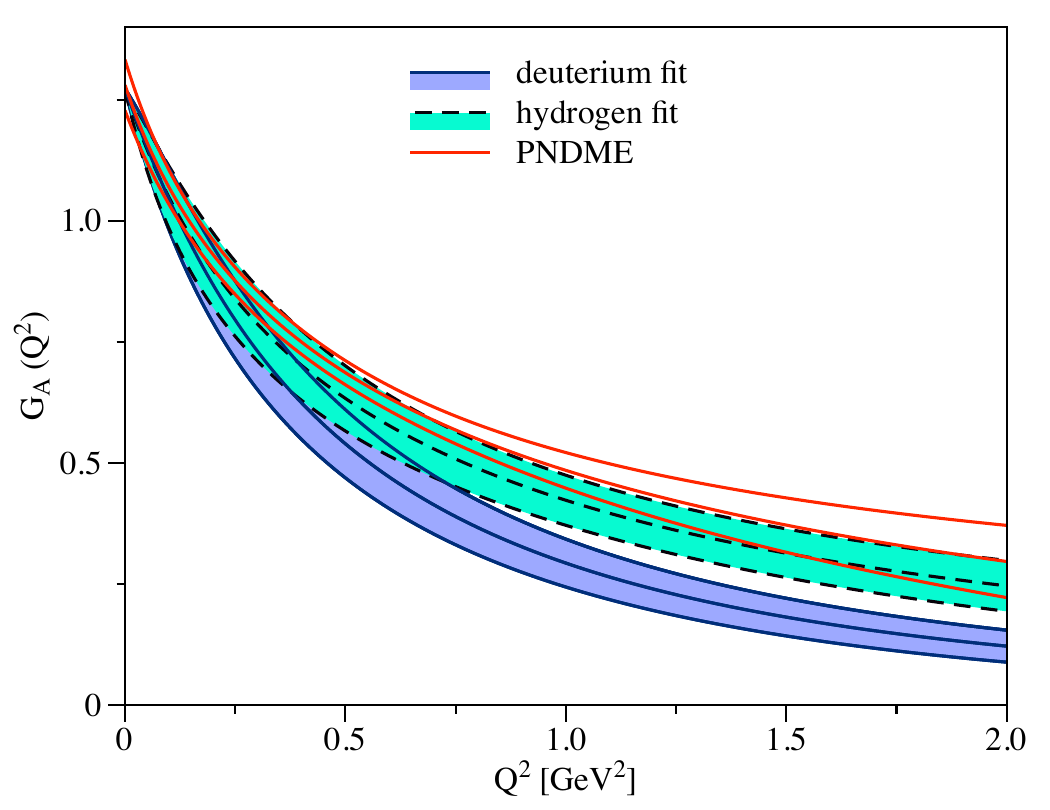}
    \hspace{0.cm}
    \includegraphics[width=0.4\textwidth]{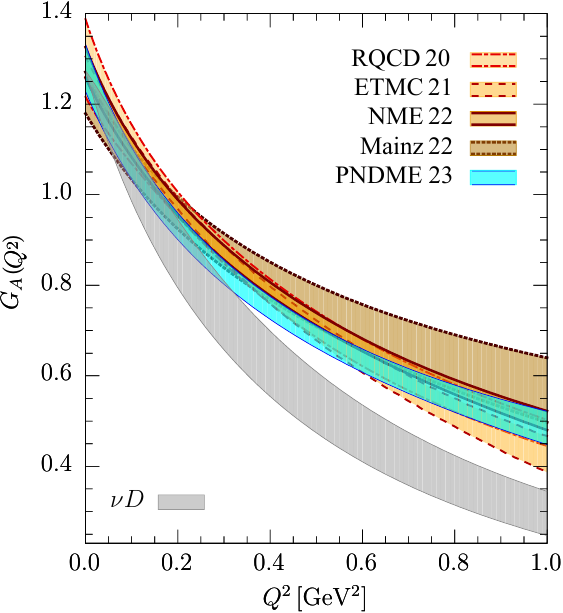}
    \caption{(Left) Comparison of the nucleon axial-vector form factor $G_A \left( Q^2 \right) = - F_A \left( Q^2 \right)$ as a function of the momentum transfer squared $Q^2$ obtained from (i) fit to the deuterium bubble-chamber data~\protect\cite{Meyer:2016oeg} shown by blue solid lines with error band; (ii) fit to recent MINERvA antineutrino-hydrogen data~\protect\cite{MINERvA:2023avz}, shown by black dashed lines and turquoise  error band; and (iii) lattice QCD result obtained by the PNDME Collaboration~\protect\cite{Jang:2023zts} shown by red solid lines {\it{without a band}}. (Right) A comparison of LQCD axial-vector form factors from various collaborations labeled RQCD 19~\protect\cite{RQCD:2019jai}, ETMC 21~\protect\cite{Alexandrou:2020okk}, NME 22~\protect\cite{Park:2021ypf}, Mainz 22~\protect\cite{Djukanovic:2022wru}, and PNDME 23~\protect\cite{Jang:2023zts}. The $\nu$D~\protect\cite{Meyer:2016oeg} band is the same as the deuterium fit shown in the left panel.}
\label{figure:axial_form_factor}
\end{figure*}

Over the last few years, a number of LQCD collaborations have presented first-principles calculations of the isovector axial-vector form factor, and their results for $G_A(Q^2) \equiv -F_A(Q^2)$ compiled in~\cite{Jang:2023zts} are reproduced in the right panel in Fig.~\ref{figure:axial_form_factor}. They agree amongst themselves, but disagree with existing evaluations based on the deuterium data~\cite{Meyer:2022mix}. Of particular note, the results by the PNDME lattice-QCD Collaboration have \({}\lesssim 10\%\) uncertainty in the Euclidean momentum transfer squared region $0<Q^2<1~\mathrm{GeV}^2$~\cite{Jang:2023zts} and, as discussed below, disagree significantly with the fits to the deuterium data.

In this paper, we take the PNDME calculation as representative of the first-principles nucleon axial-vector form factor, and show that the cross sections obtained using it are in good agreement with the MINERvA hydrogen data within the $Q^2$ range of validity of these LQCD calculations. The salient features of the PNDME calculation, which demonstrate controls essential for all reliable LQCD analyses of the axial form factors, are~\cite{Jang:2023zts}:
\begin{itemize}
    \item High statistics: PNDME analyzed $13$ ensembles of 2+1+1-flavors of highly improved staggered quarks (HISQ) using Wilson-clover valence quarks. The statistics and range of lattice spacing, $0.057~\mathrm{fm} < a < 0.151~\mathrm{fm}$, and pion masses, $M_\pi \approx 135, 220, 310$~MeV, allowed the authors to make a careful study of the various systematics. 
    \item Removal of contributions due to excited states. All correlation functions calculated on the lattice get contributions from all excited states that couple to, and are thus created, by the interpolating operators used. This problem can be severe for nucleons especially if towers of multihadron states, starting with the $N \pi$ states that have mass gaps starting at $\approx 1200$~MeV (much smaller than the $N(1440)$ radial excitation) as $M_\pi \to 135$~MeV, make large contributions. This has been shown to be the case for the axial channel~\cite{Jang:2019vkm}. The PNDME calculation includes a detailed analysis to remove contributions of such excited states.
    \item Satisfying, to within the expected size of discretization errors, the partially conserved axial current (PCAC) relation between the three form factors, axial $F_A(Q^2)$, induced pseudoscalar $F_P(Q^2)$, and pseudoscalar $G_P(Q^2)$, obtained after removing contributions from $N \pi$ excited states. Since the lattice correlation functions automatically satisfy the PCAC relation, this is a check of the decomposition into form factors that relies on the absence of transition matrix elements to excited states. It is a necessary requirement that must be satisfied by all LQCD calculations of the three form factors. Note that PNDME paper uses the notation $G_A(Q^2) \equiv - F_A(Q^2)$ and ${\widetilde G}_P(Q^2) \equiv - F_P(Q^2)/2$.
    \item The data for $F_A(Q^2)|_{\{a, M_\pi,M_\pi L\}}$ obtained at discrete values of $Q^2$ on each of the thirteen ensembles is well-fitted using the model-independent $z$-expansion. The lattice size $L$ is in units of $M_\pi$.
    \item Extrapolation of the thirteen $F_A(Q^2)|_{\{a, M_\pi,M_\pi L\}}$ to get the form factor at the physical point, $a = 0$ and $M_\pi = 135$~MeV,  is carried out for eleven equally spaced values of $Q^2$ between 0--1~GeV${}^2$ using the leading-order corrections in ${\{a, M_\pi,M_\pi L\}}$. This full analysis is done within a single overall bootstrap process and the reasonableness of the resulting error estimates are discussed. The finite-volume artifacts are found to be small for $M_\pi L \gtrsim 4$, which holds for all but two ensembles.
    \item All fits to $F_A(Q^2)$ are presented using the $z^2$ truncation of the $z$-expansion. Results with $z^3$ truncation give essentially the same values, indicating convergence. The $z^2$ results were chosen to avoid overparameterization as defined by the Akaike Information Criterion (AIC)~\cite{Akaike:1100705}.
\end{itemize}

Raw lattice data with reliable error estimates are available at discrete values of $Q^2$ over a limited range of momentum transfer, $0<Q^2 \lesssim 1~\mathrm{GeV}^2$. As shown below, for the calculation of the cross section outside  this range, a robust parameterization of the form factor is needed to connect to the $1/Q^4$ behavior (with possible logarithmic corrections) expected at large $Q^2$~\cite{Lepage:1980fj,Chernyak:1983ej}. This is typically done by enforcing sum rules~\cite{Lee:2015jqa}. This has not been done in the PNDME analysis~\cite{Jang:2023zts}.  It is, therefore, reasonable to make comparisons of the lattice and the experimental determinations for the (anti)neutrino-nucleon charged-current elastic cross sections for differential distributions only at $Q^2 \lesssim Q_\mathrm{max}^2 \approx 1~\mathrm{GeV}^2$. For total cross sections with (anti)neutrino energy $E_\nu \lesssim M \left(\tau_\mathrm{max} + r^2_\ell \right) \left( 1+ \sqrt{1 + 1/\tau_\mathrm{max}}\right) \approx 0.84~\mathrm{GeV}$, where $r_\ell = \frac{m_\ell}{2 M}$, $\tau_\mathrm{max}=\frac{Q_\mathrm{max}^2}{4M^2}$, $M$ is the nucleon mass and $m_\ell$ is the charged lepton mass, the kinematically-allowed phase space is restricted to momentum transfers $Q^2 \lesssim Q_\mathrm{max}^2 \approx 1~\mathrm{GeV}^2$.

We present two analyses: the main one is a direct comparison with the experimental results in Ref.~\cite{MINERvA:2023avz}. The second explains, using the total cross sections versus the (anti)neutrino energy $E_\nu$, why it will be important for DUNE ($E_\nu$ peaked at 2-3~GeV) to determine $F_A$ for $Q^2 > 1$~GeV${}^2$. The reader should, however, keep in mind two caveats in our analysis: first, we take the results of the axial-vector form factors at face value, i.e., we do not address issues of possible unresolved systematics in their extraction from either experimental data or from LQCD calculations. Second, total cross sections versus $E_\nu$ are not directly measured experimentally due to uncertainty in the reconstruction of the neutrino energy.

To evaluate the (anti)neutrino-nucleon charged-current elastic cross sections, we exploit the decomposition of the unpolarized differential cross section in terms of the structure-dependent $A,~B$, and~$C$ functions~\cite{LlewellynSmith:1971uhs}:
\begin{widetext}
\begin{equation}
\frac{d\sigma}{dQ^2} (E_\nu, Q^2) = \frac{\mathrm{G}_\mathrm{F}^2 |V_{ud}|^2}{2\pi} \frac{M^2}{E_\nu^2} \left[ \left( \tau + r_\ell^2 \right)A(\nu,~Q^2) - \frac{\nu}{M^2} B(\nu,~Q^2) + \frac{\nu^2}{M^4} \frac{C(\nu,~Q^2)}{1+ \tau} \right] \,, \label{eq:xsection_CCQE}
\end{equation}
\end{widetext}
with the kinematic variables $\nu = E_\nu/M - \tau - r^2_\ell$ and $\tau = \frac{Q^2}{4M^2}$, the CKM matrix element $V_{ud}$, and the Fermi coupling constant $\mathrm{G}_\mathrm{F}$. At tree level, the structure-dependent parameters $A$, $B$, and $C$ are expressed in terms of the nucleon electric $G^V_E$, magnetic $G^V_M$, axial $F_A$, and induced pseudoscalar $F_P$ isovector form factors as
\begin{align}
A &= \tau  \left( G^V_M \right)^2 - \left( G^V_E \right)^2 + \left( 1+ \tau \right) F^2_A  \nonumber \\
& -  r_\ell^2 \left[ \left( G^V_M \right)^2 + F^2_A + 4 F_P \left( F_A - \tau F_P \right) \right] \,, \\
B &=  4 \eta \tau  G^V_M F_A \,,  \\
C &= \tau \left( G^V_M \right)^2 + \left( G^V_E \right)^2 + (1 + \tau) F^2_A,
\end{align}
with $\eta = + 1$ for neutrino scattering and $\eta = - 1$ for antineutrino scattering. For deuterium-based calculation, we take the electromagnetic vector form factors from Ref.~\cite{Borah:2020gte} and the axial-vector form factor from Ref.~\cite{Meyer:2016oeg} as default fits to the data below $Q^2 < 1~\mathrm{GeV}^2$. Extending the fits to  $Q^2 < 3~\mathrm{GeV}^2$ using the same parameterization does not significantly change the results in this paper. For total cross sections, we integrate over the kinematically-allowed region of the momentum transfer $ Q^2_- \le Q^2 \le Q^2_+ $
\begin{align} \label{eq:range_Q2}
Q^2_\pm &= \frac{2 M E^2_\nu}{M+2 E_\nu} - 4M^2 \frac{M+E_\nu}{M+2 E_\nu} r_\ell^2 \nonumber \\
&\pm \frac{4 M^2 E_\nu}{M+2 E_\nu} \sqrt{\left( \frac{E_\nu}{2M}- r_\ell^2 \right)^2- r_\ell^2}.
\end{align}

\begin{figure*}     
    \centering
    \includegraphics[height=0.371\textwidth]{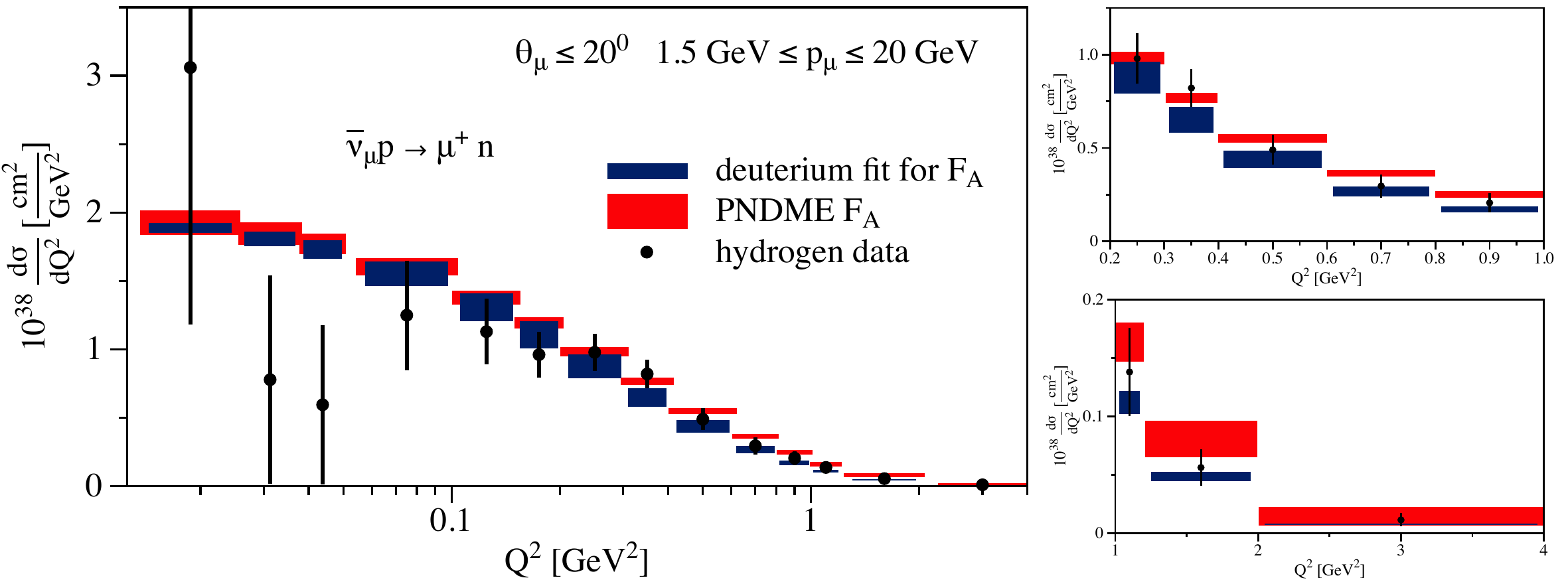}
    \caption{Antineutrino-hydrogen charged-current elastic cross-section data of MINERvA Collaboration~\cite{MINERvA:2023avz} is compared in the lower figure with theoretical prediction based on the vector form factor fit of Ref.~\cite{Borah:2020gte} and axial-vector form factor fit of Ref.~\cite{Meyer:2016oeg}, shown by the shorter dark blue bins, and with prediction based on the PNDME LQCD axial-vector form factor in Ref.~\cite{Jang:2023zts}, shown by the red bins. Kinematic cuts in the MINERvA measurement are placed on the muon scattering angle $\theta_\mu \le 20^0$ and momentum $1.5~\mathrm{GeV} \le p_\mu \le 20~\mathrm{GeV}$. The thickness of the bin size in the panels represents the error, and the fifteenth bin is not distinguished. The two right panels zoom into the region $0.2~\mathrm{GeV}^2 \lesssim Q^2 \lesssim 1~\mathrm{GeV}^2$ where LQCD predictions have the smallest errors, and the region $Q^2 \gtrsim 1~\mathrm{GeV}^2$ where errors in LQCD calculations become large.} \label{figure:MINERvA_data}
\end{figure*}

A direct comparison of the MINERvA measurement~\cite{MINERvA:2023avz,Irani:2023lol} with predictions starting with Eq.~(\ref{eq:xsection_CCQE}) for the antineutrino-hydrogen elastic differential cross sections, averaging over the incoming NUMI flux, and putting cuts on the recoil muon scattering angle $\theta_\mu \le 20^0$ and momentum $1.5~\mathrm{GeV} \le p_\mu \le 20~\mathrm{GeV}$ is shown in Fig.~\ref{figure:MINERvA_data} using the axial form factors from (i) the bubble-chamber data~\cite{Meyer:2016oeg}, and (ii) LQCD~\cite{Jang:2023zts}. We identify three regions of the momentum transfer with different significance of form factors from LQCD and existing measurements. At low momentum transfers $Q^2 \lesssim 0.2~\mathrm{GeV}^2$, LQCD predictions and fits to the deuterium bubble-chamber data are in good agreement. In this region, the experimental errors in the measurement on hydrogen by MINERvA are large, whereas the errors in the deuterium bubble chamber data are smaller, and therefore provide a benchmark. However, we remind the reader of the unresolved uncertainty due to the use of models for nuclear corrections in the extraction of form factors from the deuterium data as discussed in Ref.~\cite{Meyer:2016oeg}.

Looking ahead, calculations of $F_A$ using the current LQCD methodology will improve rapidly in this region as more simulations are done closer to $M_\pi=135$~MeV, $a \to 0$ and on larger volumes. Over time, we expect lattice results in this region to be well-characterized by the axial charge~\cite{Cirigliano:2022hob}, (precisely measured already), the axial charge radius and fits with a low-order $z$-expansion.

In the second region of momentum transfer, $0.2~\mathrm{GeV}^2 \lesssim Q^2 \lesssim 1~\mathrm{GeV}^2$, the axial-vector form factor from LQCD leads to the smallest errors and the predicted differential cross section lies above the hydrogen and deuterium values. The difference between lattice and deuterium values is due to the smaller $F_A$ from the latter as illustrated in Fig.~(20) of Ref.~\cite{Jang:2023zts}, which is reproduced in Fig.~\ref{figure:axial_form_factor} (right). Assuming no new deuterium data, no further checks against it are anticipated.  Future improvements in both the hydrogen data and lattice calculations will provide robust cross-checks in this region.

In the third region with momentum transfers $Q^2 \gtrsim 1~\mathrm{GeV}^2$, current LQCD data have large statistical errors and systematic uncertainties due to discretization errors and removing excited state contributions. It is unlikely that LQCD data in this region, coming mostly from simulations with $M_\pi \gtrsim 300$~MeV~\cite{Jang:2023zts}, will improve anytime soon. Without precise data in this region from simulations with $M_\pi \sim 135$~MeV, imposing the asymptotic $1/Q^4$ behavior using sum rules will be weighted heavily by data at $Q^2 \lesssim 0.5~\mathrm{GeV}^2$. This will result in an inherent uncertainty in lattice estimates of $F_A$ and loss of predictive power as discussed next. We, therefore, anticipate that improvements in MINERvA and follow on experiments will provide the best results in this region.

To determine the significance of the difference between the MINERvA differential cross sections and the predictions using the form factors extracted from either LQCD or the deuterium data, a \(\chi^2\) test was performed. For this, we used fifteen \(Q^2\) bins (with $Q^2 \lesssim 1~\mathrm{GeV}^2$) and the full covariance matrices. In determining the predictions using the form factors, we ignored the flux uncertainties. Comparing MINERvA-LQCD (MINERvA-deuterium), we found \(\chi^2=11\) (\(12\)), respectively, for the $15$ degrees of freedom, showing that both differences are statistically insignificant. Performing a similar comparison between the LQCD and deuterium predictions, we encountered a singular covariance matrix since both the form factors were obtained from a few-parameter fit. We examined four cases---keeping 3--6 degrees of freedom and in each dropped the small eigenvalues of the combined covariance matrix---and found the same \(\chi^2/{\rm d.o.f} \approx 6\). This implies $\approx 2.5\sigma$ tension between the LQCD and fits to the deuterium data.

\begin{figure*}   
    \centering
    \includegraphics[width=0.99\textwidth]{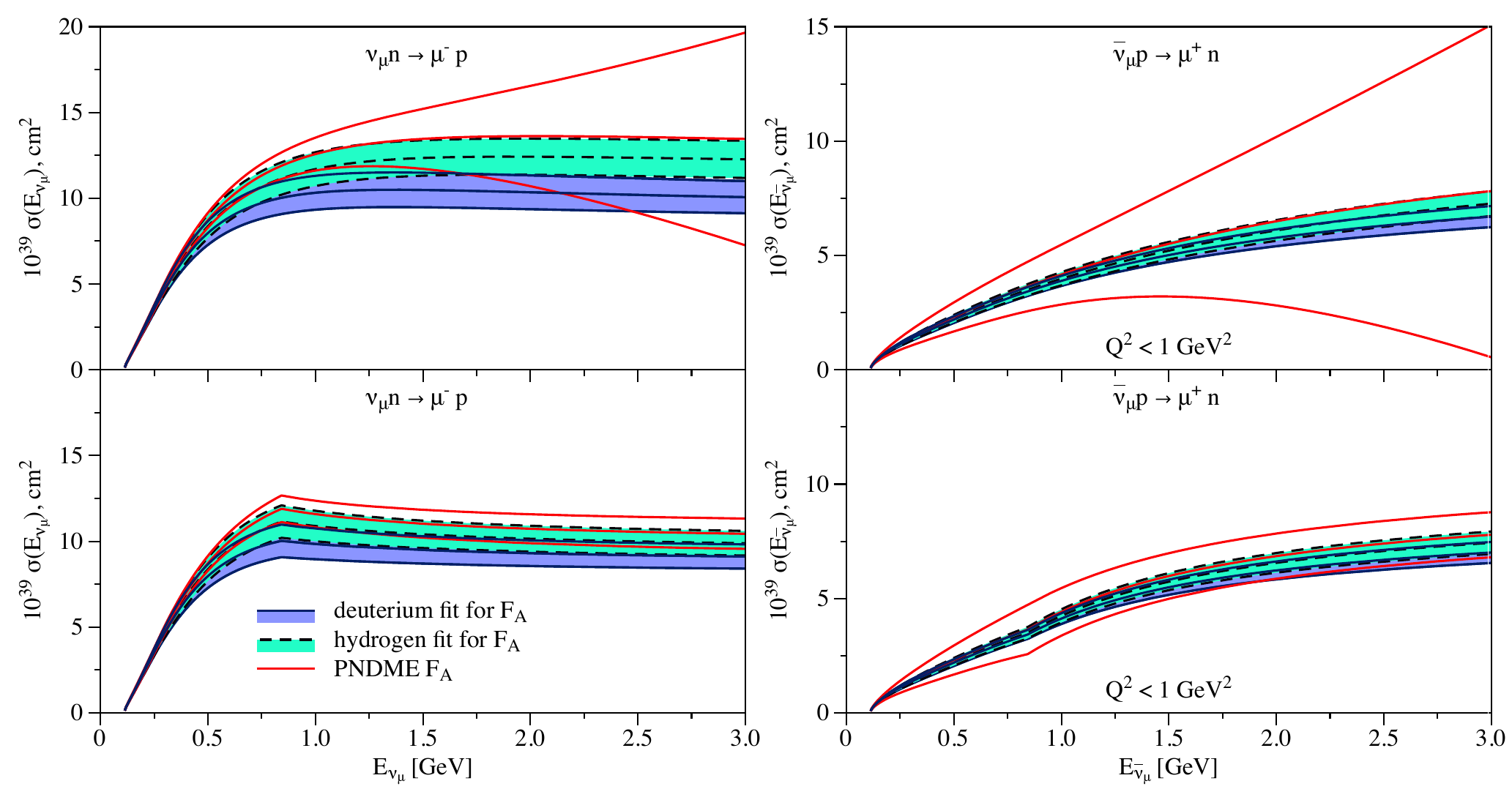}
    \caption{(Top) Neutrino-neutron and antineutrino-proton total charged-current elastic cross sections are shown versus the (anti)neutrino energy $E_{\nu}$ in the left and right panels, respectively. The prediction based on the deuterium bubble-chamber data is shown by the blue solid line and error band. The fit to recent MINERvA antineutrino-hydrogen data (labeled hydrogen fit) is shown by black dashed lines and turquoise error band. These are compared with the calculated result (red solid lines) using the PNDME axial-vector form factor and integrated over the full kinematic range. (Bottom) Same but including contributions only from the momentum transfers below $Q^2 \lesssim Q_\mathrm{max}^2 \approx 1~\mathrm{GeV}^2$, i.e., $F_A(Q^2)$ is set to zero for $Q^2 > 1~\mathrm{GeV}^2$.}\label{figure:inclusive_cross_sections_all}
\end{figure*}

Keeping in mind the second caveat stated above, we nevertheless show, in Fig.~\ref{figure:inclusive_cross_sections_all} (top), results for total neutrino-neutron and antineutrino-proton charged-current elastic cross sections based on the fits to the experimental data~\cite{Meyer:2016oeg} and the lattice determinations after the integration of Eq.~\eqref{eq:xsection_CCQE} over the $Q^2$ range in Eq.~(\ref{eq:range_Q2}). In the region $E_{\nu_\mu} \lesssim 0.84$~GeV, there is reasonable agreement ($1$--$2\sigma$) with the ordering PNDME $>$ hydrogen $>$ deuterium in accord with the pattern in $F_A$ shown in Fig.~\ref{figure:axial_form_factor}. For $E_{\nu_\mu} \gtrsim 0.84$~GeV, one goes beyond the applicability of the lattice data~\cite{Jang:2023zts}, and its predictive power fails. The right panel illustrates that the uncertainties in the antineutrino-proton cross sections using PNDME result are even larger, which can be traced to a close-to-singular structure of the covariance matrix for the axial-vector form factor reported in Ref.~\cite{Jang:2023zts}. The possibility that the range of validity of the LQCD-based predictions can be enlarged by imposing the asymptotic $1/Q^4$ behavior through sum rules needs to be checked.

Figure~\ref{figure:inclusive_cross_sections_all} (bottom) shows the same analysis but with $F_A(Q^2)$ set to zero for $Q^2> 1$~GeV${}^2$. Now the results for neutrino cross sections agree for the full range of $E_{\nu_\mu}$. The LQCD result for antineutrinos continues to show larger uncertainty for the reason mentioned above.

To summarize, we have compared (anti)neutrino-nucleon charged-current elastic cross sections based on fits to the well-known deuterium bubble-chamber data and new measurements by the MINERvA Collaboration with a theoretical analysis using a representative LQCD calculation of the nucleon axial-vector form factor. We have identified three regions of $Q^2$ in which to assess the strengths and weaknesses of experimental measurements and lattice calculations. We anticipate that LQCD will provide the best estimates for the axial-vector form factor for $Q^2 \lesssim 0.2~\mathrm{GeV}^2$. The reason is that with the current LQCD method, as the lattice size is increased at fixed $a$ or as $a$ is decreased, the value of $Q^2_{\rm max}$ at which data with good statistical precision can be obtained shrinks to $Q^2_{\max} \lesssim 0.2$~GeV${}^2$. Over the region $0.2~\mathrm{GeV}^2 \lesssim Q^2 \lesssim 1~\mathrm{GeV}^2$, both LQCD and antineutrino on hydrogen experiments will provide increasingly precise data that will lead to growing confidence in both. For $Q^2 > 1~\mathrm{GeV}^2$, experimental measurements are the best near-term option since novel lattice methodology is needed to control the systematics that grow with $Q^2$.

We anticipate significant progress in LQCD data over the next five years due to increase in both statistics that will allow control over excited states contributions through the inclusion of three or more states in the spectral decomposition of correlation functions~\cite{Jang:2019vkm,Jang:2023zts}, and the values of $\{a,M_\pi\}$ at which simulations are done, which will improve the chiral-continuum fits used to remove the associated systematics. However, to reach percent-level accuracy, novel methods~\cite{Bali:2016lva,Gao:2021xsm} are needed to remove both excited-state contributions and discretization errors with requisite precision. \looseness-1

\section*{Acknowledgments}
We acknowledge useful discussions with Aaron Meyer, Tejin Cai, Deborah Harris, Laura Fields, and Kevin McFarland. This work is supported by the US Department of Energy through the Los Alamos National Laboratory. Los Alamos National Laboratory is operated by Triad National Security, LLC, for the National Nuclear Security Administration of U.S. Department of Energy (Contract No. 89233218CNA000001). OT was supported by LANL’s Laboratory Directed Research and Development (LDRD) program under projects 20210968PRD4 and 20210190ER, TB and RG were supported by DOE HEP under Contract No. DE-AC52-06NA25396 and LDRD project number 20210041DR. Mathematica~\cite{Mathematica} and DataGraph~\cite{JSSv047s02} were used in this work.

\bibliographystyle{apsrev4-2}

\bibliography{PNDME}

\end{document}